\documentclass{article}[12pt]
\usepackage{amsmath,amssymb}
\usepackage{graphicx}%
\usepackage{xcolor}
\usepackage{verbatim}
\usepackage{multirow}

\newtheorem{stat}{Statement}

\newcommand{\Mod}[1]{\mathbf{#1}}
\newcommand{\ModTilde}[1]{\tilde{\mathbf{#1}}}

\newcommand{\TwoFigs}[4]{%
\begin{flushleft}
\begin{tabular}{cc}
\parbox{7cm}{\centerline{\includegraphics[width=7cm]{#1}}}  & \parbox{7cm}{\centerline{\includegraphics[width=7cm]{#2}}}  \\
\parbox{7cm}{\vspace{7pt}\refstepcounter{figure}Fig. \thefigure.\quad #3\vfill} & \parbox{7cm}{\vspace{7pt}\refstepcounter{figure}Fig. \thefigure.\quad #4\vfill}\\
\end{tabular}
\end{flushleft}
\vspace{7pt}
}

\newcounter{strochka}

\newcounter{spisok}
\begin{document}

\begin{center}
{\bf \Large Yu.G. Ignat'ev\footnote{Institute of Physics, Kazan Federal University, Kremlyovskaya str., 35, Kazan, 420008, Russia}\footnote{Email: yurii.ignatev.1947@yandex.ru} }\\[12pt]
{\bf \Large Similarity of cosmological models and its application\newline to the analysis of cosmological evolution} \\[12pt]
\end{center}

\abstract{Scale transformations of cosmological models based on a statistical system of degenerate fermions with a scalar Higgs interaction are studied. The similarity properties of cosmological models are revealed under the scale transformation of their fundamental parameters. The laws of transformation of the coordinates of singular points and eigenvalues of the characteristic matrix of the dynamical system of the cosmological model under its scale transformations are established. With the help of the previously studied dynamical system of scalarly charged fermions is transformed to new variables and modified to a dynamical system with a nondegenerate characteristic matrix and its nondegenerate branch, the singular points and eigenvalues of the characteristic matrix are found, which coincide with the corresponding values for the vacuum field model. Examples of numerical simulation of such cosmological models are given.\\

}


%
\section*{Introduction}
Methods of similarity theory, as well as dimensional analysis of dynamic systems \cite{Klain} have long been successfully used in mechanics, hydro- and gas dynamics \cite{Sedov}, as well as in astrophysics and cosmology \cite{Dibai}. These methods make it possible to extend the results of research to other models and are especially valuable in the study of complex, essentially nonlinear dynamic systems, when the use of numerical modeling methods becomes mandatory. Revealing the laws of similarity of such dynamical systems makes it possible to extend the results of numerical integration to models with other parameters, thereby providing the possibility of a comprehensive numerical-analytical study of a class of models.

Methods of similarity theory and dimensional analysis become especially effective in the study of cosmological models based, in turn, on various field-theoretic models containing fundamental constants and parameters that are often not defined at the time of the study. In \cite{TMF_21}, on the basis of microscopic dynamics, a macroscopic model of the Universe was formulated, based on statistical systems of fermions with scalar charges, classical and phantom, with the Higgs interaction potential, and in \cite{Ignat21_1} -- \cite{Ignat22_1} various versions of this model were constructed and studied. Subsequently, these models were investigated for stability with respect to longitudinal plane-wave gravitational perturbations \cite{Yu_Unst_GC22_1} -- \cite{Yu_Unst_GC23_1}. In these works, the short-wavelength scalar-gravitational instability of a homogeneous cosmological model, which is fundamentally different from the previously studied hydrodynamic and gaseous gravitational instability, was revealed and studied. In the same works, the fundamental possibility of supermassive black holes in the early Universe was shown using the mechanism of scalar-gravitational instability. In \cite{YuTMF_23}, the evolution of spherical gravitational perturbations in a medium of scalarly charged fermions with the Higgs interaction was studied, in particular, the evolution of localized perturbations without wavelength limitation.

Studies on the scalar-gravitational stability of the cosmological medium of degenerate scalarly charged fermions, in particular, revealed a close connection between the singular points of the vacuum-field cosmological model and the appearance of unstable phases in the model with charged fermions. This revealed connection makes it necessary to carry out a more detailed qualitative study of the dynamical systems of cosmological models based on statistical systems of scalarly charged fermions. To study the general regularities for these models, the methods of similarity theory also became necessary. This article is devoted to these questions. Note that the general similarity property for self-gravitating systems with scalar Higgs interaction was formulated in the cited work \cite{YuTMF_23}.
\section{Cosmological system of fermions with scalar interaction}
Let us briefly formulate the main provisions of the macroscopic theory\footnote{In \cite{TMF_21} it is shown how this theory is obtained from microscopic dynamics.} for a cosmological model based on a one-component degenerate statistical system of scalarly charged fermions and a scalar Higgs field $\Phi$.
%
\subsection{General Model Equations}
The Lagrange function $L_s$ of the scalar Higgs field is\footnote{Here and below, Latin letters run over $\overline{1,4}$, Greek -- $\overline{1,3}$.}
\begin{eqnarray} \label{L_s}
L_s=\frac{1}{16\pi}(g^{ik} \Phi_{,i} \Phi_{,k} -2V(\Phi)),
\end{eqnarray}
where
\begin{eqnarray}
\label{Higgs}
V(\Phi)=-\frac{\alpha}{4} \left(\Phi^{2} -\frac{m_s^{2}}{\alpha}\right)^{2}
\end{eqnarray}
is the potential energy of the scalar field, $\alpha$ is the self-action constant, $m_s$ is the mass of the quanta of the scalar field. Energy-momentum tensors of scalar fields relative to the Lagrange function \eqref{L_s}, $S^i_{k}$, and \emph{equilibrium} statistical system, $T^i_{k}$, are
\begin{eqnarray}\label{T_s}
S^i_{k} =\frac{1}{16\pi}\bigl(2\Phi^{,i}\Phi_{,k}- \delta^i_k\Phi_{,j} \Phi^{,j}+2V(\Phi)\delta^i_k \bigr);
\end{eqnarray}
\begin{equation}\label{T_p}
T^i_{k}=(\varepsilon_p+p_p)u^i u_k-\delta^i_k p_p,
\end{equation}
where $u^i$ is the velocity vector of the statistical system and $\varepsilon_p$, $p_p$ are the energy density and pressure of the statistical system.

Einstein's equations for the ``scalar field+particles'' system are:
\begin{equation}\label{Eq_Einst_G}
R^i_k-\frac{1}{2}\delta^i_k R=8\pi (T^i_k+S^i_k) + \delta^i_k \Lambda_0,
\end{equation}
where $\Lambda_0$ is the seed value of the cosmological constant, related to its observed value $\Lambda$, obtained by removing the constant term in the potential energy, by the relation:
\begin{equation}\label{lambda0->Lambda}
\Lambda=\Lambda_0-\frac{m_s^4}{4\alpha}.
\end{equation}

The macroscopic consequences of the kinetic theory are the transfer equations \cite{TMF_21}, including the law of conservation of some vector current corresponding to the microscopic law of conservation of some fundamental charge %
\begin{equation}\label{1}
\nabla_i q n^i=0,
\end{equation}
as well as the laws of conservation of energy - momentum of the statistical system:
\begin{equation}\label{2}
\nabla _{k} T_{p}^{ik} -\sigma\nabla^{i} \Phi =0,
\end{equation}
where $\sigma$ is the density of scalar charges with respect to the field $\Phi$ \cite{TMF_21}. The \eqref{2} equations are equivalent to the equations of ideal hydrodynamics
\begin{eqnarray}\label{2a}
(\varepsilon_p+p_p)u^i_{~,k}u^k=(g^{ik}-u^iu^k)(p_{p,k}+\sigma\Phi_{,k});\\
\label{2b}
\nabla_k[(\varepsilon_p+p_p)u^k]=u^k(p_{p,k}+\sigma\Phi_{,k}),
\end{eqnarray}
and the fundamental charge conservation laws \eqref{1}:
\begin{equation}\label{2c}
\nabla_k \rho u^k=0,
\end{equation}
where $\rho\equiv q n$ is \emph{the kinematic density of the scalar charge}.

Macroscopic scalars for a one-component statistical system of degenerate fermions take the form:
\begin{eqnarray}
\label{2_3c}
n=\frac{1}{\pi^2}\pi_f^3;\; p_p  =\displaystyle \frac{e^4\Phi^4}{24\pi^2}(F_2(\psi)-4F_1(\psi));\\
\label{2_3a_2}
 \sigma=\frac{e^4 \Phi^3}{2\pi^2}F_1(\psi);   \; \varepsilon_p=\frac{e^4 \Phi^4}{8\pi^2}F_2(\psi),
\end{eqnarray}
where $\pi_f$ is the Fermi momentum, $\sigma$ is the density of scalar charges $e$ and
\begin{equation}\label{psi}
\psi=\frac{\pi_f}{|e\Phi|}
\end{equation}
and the functions $F_1(\psi)$ and $F_2(\psi)$ are introduced:
\begin{eqnarray}\label{F_1}
F_1(\psi)=\psi\sqrt{1+\psi^2}-\ln(\psi+\sqrt{1+\psi^2});\nonumber\\
\label{F_2}
F_2(\psi)=\psi\sqrt{1+\psi^2}(1+2\psi^2)-\ln(\psi+\sqrt{1+\psi^2}).\nonumber
\end{eqnarray}
Wherein, the scalar field equation for a system of scalarly charged degenerate fermions is obtained as a consequence of the transport equations
\begin{eqnarray}\label{Box(Phi)=sigma_z}
\Box \Phi + m_s^2\Phi-\alpha\Phi^3 =-8\pi\sigma\equiv-\frac{4e^4\Phi^3}{\pi} F_1(\psi).
\end{eqnarray}
Thus, the complete system of equations of the $\Mod{M}$ mathematical model of a system of scalarly charged fermions consists of Einstein's equations \eqref{Eq_Einst_G}, hydrodynamic equations \eqref{2} and the scalar field equation \eqref{Box(Phi)=sigma_z} together with definitions of the sources: scalar field energy-momentum tensors \eqref{T_s}, fermionic component, \eqref{T_p}, and the scalar charge density, \eqref{2_3a_2}, as well as the fermion energy density \eqref{2_3c} and their pressure \eqref{2_3a_2}. As can be seen from the equations of this system and the definition of its coefficients, the solution of the Cauchy problem for this system of equations for given \emph{fundamental parameters}
\begin{equation}\label{p_0}
\Mod{p}=[[\alpha,m_s,e],\Lambda]
\end{equation}
are completely determined by the corresponding initial conditions for the metric functions $g_{ik}(x^j)$, the potential $\Phi(x^j)$, the velocity vector $u^i(x^j)$, and the Fermi momentum $\pi_f(x^j)$. This complete system of equations, together with the definitions of the functions contained in them, given by the initial conditions on the Cauchy hypersurface and a given set of fundamental parameters $\Mod{p}$, will be referred to as the mathematical model $\Mod{M}$ of a self-gravitating statistical system of degenerate scalarly charged fermions with the Higgs interaction.

\subsection{The similarity property of a mathematical model}
In \cite{YuTMF_23} the following similarity property of the considered dynamical system is proved.\\[6pt]

\begin{stat}\label{stat1}
$\blacksquare$ The complete system of equations of the mathematical model $\Mod{M}$ is invariant with respect to simultaneous scaling transformations of fundamental parameters $\Mod{P}$ \eqref{trans_param}, coordinates and Fermi momentum \eqref{trans_x} of the mathematical model
\begin{eqnarray}\label{trans_param}
\mathcal{S}_k(\Mod{M}): &  \alpha=k^2\tilde{\alpha},\; m_s=k\tilde{m}_s;\; e=\sqrt{k}\tilde{e};\;\Lambda= k^2\tilde\Lambda;&\\
\label{trans_x}
& x^i= k^{-1}\tilde{x}^i,\quad \pi_f= \sqrt{k}\tilde{\pi}_f, & (k=\mathrm{Const}>0).
\end{eqnarray}
that is, under scaling transformations \eqref{trans_param} -- \eqref{trans_x} and the corresponding transformation of the initial conditions, the solutions of the equations of the original model $\Mod{M}$ and the scaling transformed model $\ModTilde{M}$ coincide:
\begin{eqnarray}\label{trans_eqs}
\Phi(x)=\tilde{\Phi}(\tilde{x});\; g_{ik}(x)= \tilde{g}_{ik}(\tilde{x});\; u^i(x)= \tilde{u}^i(\tilde{x}).& \blacksquare
\end{eqnarray}
\end{stat}
\vspace{12pt}
The similarity property of a mathematical model allows extending a solution with a given set of fundamental parameters to the case of other values of fundamental parameters. This is practically important in the numerical integration of the model equations in the case of very small or very large values of the parameters and over large intervals of coordinate values.

Under scaling transformations \eqref{trans_param} -- \eqref{trans_x} both parts of the equations \eqref{Eq_Einst_G}, \eqref{2} and \eqref{Box(Phi)=sigma_z} are multiplied by $k^2$, and the scalars and tensors introduced above change according to the laws:
\begin{eqnarray}\label{trans_scalar}
\psi=\tilde{\psi},\; \sigma= k^2\tilde{\sigma};\;V(\Phi)= k^2 \tilde{V}(\tilde{\Phi});\nonumber\\
p_p= k^2 \tilde{p}_p;\; \varepsilon_p= k^2 \tilde{\varepsilon}_p;
S^i_k= k^2 \tilde{S}^i_k;\; T^i_k= k^2 \tilde{T}^i_k.
\end{eqnarray}
\subsection{Cosmological model equations}
In the case of a spatially flat Friedmann metric
\begin{eqnarray}\label{ds_0}
ds_0^2=dt^2-a^2(t)(dx^2+dy^2+dz^2),
\end{eqnarray}
and homogeneous isotropic distribution of matter $\Phi=\Phi(t);\; \pi_f=\pi_f(t);\; u^i=\delta^i_4$ the energy-momentum tensor of a scalar field takes the form of the energy-momentum tensor of an ideal isotropic fluid:
\begin{equation} \label{MET_s}
S^{ik} =(\varepsilon_s +p_{s} )u^{i} u^{k} -p_s g^{ik} ,
\end{equation}
where:
\begin{eqnarray}\label{Es-Ps}
\varepsilon_s=\frac{1}{8\pi}\biggl(\frac{\dot{\Phi}^2}{2}+V(\Phi)\biggr);\nonumber\\
p_{s}=\frac{1}{8\pi}\biggl(\frac{\dot{\Phi}^2}{2}-V(\Phi)\biggr).
\end{eqnarray}
In this case, the material equations \eqref{2} -- \eqref{2a} are exactly integrated \cite{TMF_21}:
\begin{equation}\label{aP0}
 a\pi_f=\mathrm{Const},
\end{equation}
as a result, the function $\psi$ \eqref{psi} is defined in terms of the functions $a(t)$ and $\Phi(t)$:
\begin{equation}\label{psi(t)}
\psi=\frac{\pi_0}{|e\Phi|}\mathrm{e}^{-\xi}, \quad (\pi_0=\pi_f(0)),
\end{equation}
where we put here and hereafter
\begin{equation}\label{a-xi}
\xi=\ln a,\quad \xi_0\equiv \xi(0)=0.
\end{equation}

Thus, the system of equations \eqref{Eq_Einst_G}, \eqref{1}, \eqref{2} and \eqref{Box(Phi)=sigma_z} reduces to the autonomous dynamic system \cite{TMF_21}\footnote{$H(t)$ -- Hubble parameter}:
\begin{eqnarray}\label{dot_xi-dot_Phi}
\dot{\xi}=H\; (\equiv F_1),\; \dot{\Phi}=Z\;(\equiv F_3),\\
\label{dH/dt_0}
\dot{H}=- \frac{Z^2}{2}-\frac{4}{3\pi}e_z^4\Phi^4\psi^3\sqrt{1+\psi^2}\;(\equiv F_2),\\
\label{dZ/dt}
\dot{Z}=-3HZ-m_s^2\Phi +\nonumber\\
\Phi^3\biggl(\alpha-\frac{4e^4}{\pi}F_1(\psi)\biggr)\;(\equiv F_4),
\end{eqnarray}
and the Einstein equation for the $^4_4$ component becomes the first integral of this system:
\begin{eqnarray}\label{Surf_Einst1_0}
3H^2-\Lambda-\frac{Z^2}{2}-\frac{m_s^2\Phi^2}{2}+\frac{\alpha\Phi^4}{4}-\frac{e^4\Phi^4}{\pi}F_2(\psi)=0.
\end{eqnarray}
The \eqref{Surf_Einst1_0} equation defines some three-dimensional hypersurface $\mathbb{S}_3$ in the four-dimensional arithmetic phase space of a dynamical system
\begin{equation}\label{R_4}
\mathbb{S}_3\subset \mathbb{R}_4=\{\xi,H,\Phi,Z\}\equiv \{x_1,x_2,x_3,x_4\},
\end{equation}
on which all phase trajectories of the dynamical system lie, i.e., specific cosmological models\footnote{Following \cite{Yu_Dima_20GC2} we will call $\mathbb{S}_3$ \emph{Einstein-Higgs hypersurface}.}. The \eqref{Surf_Einst1_0} equation determines the initial value of the Hubble parameter $H(0)\equiv H_0$ given the initial values of the remaining dynamic variables. Two symmetric solutions for the initial value of the Hubble parameter $H^\pm_0=\pm H_0$ correspond to starting from the expansion state ($+$) or from the contraction state ($-$). The autonomous system \eqref{dot_xi-dot_Phi} -- \eqref{dZ/dt} is invariant with respect to time translations $t\to t+t_0$, which allows us to choose \eqref{a-xi} ($\xi_0=0$) as the initial condition. Thus, under the condition when choosing the sign of the initial value of the Hubble parameter, only two initial values remain free: $\Phi_0$ and $Z_0$, which we will also specify by an ordered list
\begin{eqnarray}\label{Inits}
\mathbf{I}=[\Phi_0,Z_0], & (\Phi_0=\Phi(0),Z_0=Z(0)).
\end{eqnarray}
Taking into account the exact integral \eqref{aP0}, the initial value of the Fermi momentum $\pi_0$ will also be assumed to be the fundamental parameter of the cosmological model, setting further the fundamental parameters of the model $\mathbf{M}$
ordered list \cite{TMF_21}:
\begin{eqnarray}\label{Par}
\mathbf{P}=[[\alpha,m_s,e,\pi_0],\Lambda].
\end{eqnarray}
\subsection{Similarity of cosmological models}

So, consider two cosmological models: $\mathbf{M}$ with fundamental parameters $\mathbf{P}$ and initial conditions $\mathbf{I}$
and a similar model $\tilde{\mathbf{M}}$ with fundamental parameters $\tilde{\mathbf{P}}$ and initial conditions $\tilde{\mathbf{I}}$ --
\begin{eqnarray}
\label{Inits_tilde}
\tilde{\mathbf{I}}=\biggl[\Phi_0,\frac{1}{k}Z_0\biggl];\\
\label{Par_tilde}
\tilde{\mathbf{P}}=\biggl[\biggl[\frac{\alpha}{k^2},\frac{m_s}{k},\frac{e}{\sqrt{k}},\frac{\pi_0}{\sqrt{k}}\biggr],\frac{\Lambda}{k^2}\biggr].
\end{eqnarray}
Functions $f(t)=\tilde{f}(\tilde{t})$ that are invariant under the similarity transformation \eqref{trans_param} -- \eqref{trans_x} are transformed according to the rules:
\begin{equation}\label{f=f}
\tilde{f}(\tilde{t})=f\biggl(\frac{\tilde{t}}{k}\biggr).
\end{equation}
Let the solutions of the dynamical system \eqref{dot_xi-dot_Phi} -- \eqref{dZ/dt}, \eqref{Surf_Einst1_0} for the model $\mathbf{M}$ \eqref{Inits} -- \eqref{Par} be
\[\mathbf{S}(t)=[\xi(t),H(t),\Phi(t),Z(t)].\]
Then the solutions of the corresponding equations for such a model $\tilde{\mathbf{M}}$ \eqref{Inits_tilde} -- \eqref{Par_tilde} are
\begin{eqnarray}\label{Sol_tilde}
\tilde{\mathbf{S}}(t)=[\tilde{\xi}(t),\tilde{H}(t),\tilde{\Phi}(t),\tilde{Z}(t)]\equiv 
\biggl[\xi\biggl(\frac{t}{k}\biggl),\frac{1}{k}H\biggl(\frac{t}{k}\biggl),\Phi\biggl(\frac{t}{k}\biggl),\frac{1}{k}Z\biggl(\frac{t}{k}\biggl)\biggr].
\end{eqnarray}
\subsection{Transformation of the matrix of a dynamical system and its eigenvalues}
Let us find out how \eqref{trans_param} -- \eqref{trans_x} transforms under scaling transformations the eigenvalues of the matrix of the dynamical system.
The characteristic matrix $\mathbf{A}$ of the dynamical system $\mathbf{M}$ \eqref{dot_xi-dot_Phi} -- \eqref{dZ/dt} at the point $M$ (see, for example, \cite{Bogoyav}) according to the phase coordinates \eqref{R_4} is:
\begin{eqnarray}\label{A}
\mathbf{A}(M)=\left|\left|A_i^k \right|\right|\equiv \left|\left|\frac{\partial F_i}{\partial x_k}\right|\right|=
\left(\begin{array}{cccc}
0                   & 1   &   0 & 0 \\
\displaystyle\frac{\partial F_2}{\partial \xi}  & 0 & \displaystyle\frac{\partial F_2}{\partial \Phi} & -Z \\
 0 & 0 & 0 & 1 \\
\displaystyle\frac{\partial F_4}{\partial \xi} & -3Z & \displaystyle\frac{\partial F_4}{\partial \Phi} & - 3H \\
\end{array}\right).\nonumber
\end{eqnarray}

According to the law of transformation of fundamental parameters \eqref{trans_param}, coordinates and Fermi momentum, the right parts of the dynamical system \eqref{dot_xi-dot_Phi} -- \eqref{dZ/dt} of a similar dynamical system
$\tilde{\mathbf{M}}$ are obtained according to the rules:
\begin{eqnarray}\label{tilde_F=}
\tilde{F_1}=\frac{1}{k}F_1;\; \tilde{F_2}=\frac{1}{k^2}F_2;\nonumber\\
\tilde{F_3}=\frac{1}{k}F_3;\; \tilde{F_4}=\frac{1}{k^2}F_4.
\end{eqnarray}
Thus, the characteristic matrix $\tilde{\mathbf{A}}$ of the image $\tilde{\mathbf{M}}$ of the dynamical system is:
\begin{eqnarray}\label{tilde_A}
\tilde{\mathbf{A}}(\tilde{M})=\left|\left|\tilde{A}_i^k \right|\right|\equiv \left|\left|\frac{\partial \tilde{F}_i}{\partial \tilde{x}_k}\right|\right|=
\left(\begin{array}{cccc}
0                   & 1   &   0 & 0 \\
\displaystyle\frac{1}{k^2}\frac{\partial F_2}{\partial \xi}  & 0 & \displaystyle\frac{1}{k^2}\frac{\partial F_2}{\partial \Phi} &\displaystyle -\frac{1}{k}Z \\
 0 & 0 & 0 & 1 \\
\displaystyle\frac{1}{k^2}\frac{\partial F_4}{\partial \xi} & \displaystyle -\frac{3}{k}Z & \displaystyle\frac{1}{k^2}\frac{\partial F_4}{\partial \Phi} & \displaystyle -\frac{3}{k}H \\
\end{array}\right).\nonumber
\end{eqnarray}

Comparison of \eqref{A} and \eqref{tilde_A} shows that the characteristic matrix $\tilde{\mathbf{A}}$ of the image $\tilde{\mathbf{M}}$ of the dynamical system $\mathbf{M}$ is not similar to the matrix $\mathbf{A}$, which may lead us to the wrong conclusion. It must be borne in mind that this matrix is not an independent object, but is connected precisely with the characteristic equation of the qualitative theory of dynamical systems:
\begin{equation}\label{Char_eq}
\mathbf{A}\cdot \mathbf{X}=\lambda \mathbf{X},\quad \mathbf{X}=\left(\begin{array}{c}
\xi \\ H\\ \Phi\\ Z\\
\end{array}
\right).,
\end{equation}
where $\lambda$ are eigenvalues, and $X$ is a column matrix of phase coordinates of a point in the dynamical system. The corresponding equations for the image $\tilde{\mathbf{M}}$ will look like:
\begin{equation}\label{Char_eq_tilde}
\tilde{\mathbf{A}}\cdot \tilde{\mathbf{X}}=\tilde{\lambda} \tilde{\mathbf{X}},\quad \tilde{\mathbf{X}}=\left(\begin{array}{c}
\xi \\[2pt] \displaystyle\frac{H}{k}\\[4pt] \Phi\\[2pt]  \displaystyle\frac{Z}{k}\\
\end{array}
\right).
\end{equation}
Multiplying according to \eqref{Char_eq} the matrices $\mathbf{A}$ \eqref{A} and $\mathbf{X}$ and according to \eqref{Char_eq_tilde} the matrices $\tilde{\mathbf{A}}$ \eqref{tilde_A} and $\tilde{\mathbf{X}}$, it is easy to see that the resulting systems
linear homogeneous algebraic equations with respect to phase coordinates become equivalent if and only if the eigenvalues $\lambda$ and $\tilde{\lambda}$ are related by the relation $\tilde{\lambda}=\lambda/k$.

Thus, the following statement is true:
\begin{stat}\label{stat2}
$\blacksquare$ Under scaling transformations \eqref{trans_param} -- \eqref{trans_x} the eigenvalues of the characteristic matrix of the dynamical system \eqref{dot_xi-dot_Phi} -- \eqref{dZ/dt} are transformed according to the law
\begin{equation}\label{tilde_lambda=}
\tilde{\lambda}=\frac{\lambda}{k}.
\end{equation}
In this case, the coordinates of singular points are transformed, as well as arbitrary coordinates of the phase trajectory of the dynamical system, i.e., according to the law \eqref{Char_eq_tilde} (or, equivalently, according to the law \eqref{Sol_tilde}).

Due to the proportionality of the eigenvalues of such models, the nature of singular points is an invariant property of similarity.
$\blacksquare$
\end{stat}

\subsection{Vacuum - field cosmological model}
The transition to the vacuum-field cosmological model, in which there is no scalar-charged matter, is carried out by substituting $e=0$ in the system of equations \eqref{dH/dt_0} -- \eqref{Surf_Einst1_0}. As a result, we obtain a system of equations:
\begin{eqnarray}\label{dot_xi-dot_Phi_e=0}
\dot{\xi}=H;\; (\equiv F_1),\; \dot{\Phi}=Z;\; (\equiv F_3),\\
\label{dH/dt_0_e=0}
\dot{H}=- \frac{Z^2}{2};\; (\equiv F_2),\\
\label{dZ/dt_e=0}
\dot{Z}=-3HZ-m_s^2\Phi+\alpha\Phi^3,\; (\equiv F_4),
\end{eqnarray}
\begin{eqnarray}\label{Surf_Einst1_0_e=0}
3H^2-\Lambda-\frac{Z^2}{2}-\frac{m_s^2\Phi^2}{2}+\frac{\alpha\Phi^4}{4}=0.
\end{eqnarray}
This dynamical system is a special case of the general \eqref{dot_xi-dot_Phi} -- \eqref{dZ/dt} system, so it inherits all the similarity properties discussed above. On the other hand, as studies \cite{TMF_21}, \cite{Ignat21_1} -- \cite{Ignat22_1} showed, the cosmological system of scalarly charged particles inherits the behavior of vacuum-field cosmological models, and therefore the importance of studying their global properties remains. However, this dynamical system also has a fundamental difference from the dynamical system \eqref{dot_xi-dot_Phi} -- \eqref{dZ/dt} considered above -- all functions $F_i$ of this system do not depend on the scaling function $\xi(t)$. As a result, the dynamical system \eqref{dot_xi-dot_Phi_e=0} -- \eqref{dZ/dt_e=0} is reduced to an autonomous subsystem in the three-dimensional phase space $R_3=\{H,\Phi,Z\}$. Reducing the dimension of the phase space, in turn, leads to the removal of conditions on the Hubble parameter by the subsystem of dynamic equations, -- as a result, the Hubble parameter at singular points is determined from the Einstein equation \eqref{Surf_Einst1_0_e=0}.
Let us demonstrate the above by analyzing the singular points of the single-field model \eqref{dot_xi-dot_Phi_e=0} -- \eqref{Surf_Einst1_0_e=0}.
The singular points of the model are determined by the equality to zero of the right parts of the dynamic equations, i.e., their phase coordinates are determined by the system of equations:
\begin{equation}\label{Eq_sing_points_gen}
F_i(\mathbf{X})=0, \quad (i=\overline{1,4}).
\end{equation}
For the \eqref{dot_xi-dot_Phi_e=0} -- -- \eqref{dZ/dt_e=0} system, the \eqref{Eq_sing_points_gen} equations and their solutions take the form:
\begin{eqnarray}\label{Eq_sing_points_1-3}
H=0;\quad Z=0;\\
\label{Phi_sing}
-m_s^2\Phi+\alpha\Phi^3=0,\Rightarrow \nonumber\\
\Phi_0=0; \Phi_\pm=\pm\frac{m}{\sqrt{\alpha}}.
\end{eqnarray}
Substituting these solutions into the first integral \eqref{Surf_Einst1_0_e=0} leads to conditions on the value of the cosmological constant:
\begin{eqnarray}\label{Lambda0=0}
\begin{array}{lcl}
\Phi=\Phi_0&\Rightarrow &\Lambda=0;\\
\Phi=\Phi_\pm &\Rightarrow &\Lambda_0=0,\\
\end{array}
\end{eqnarray}
where $\Lambda_0$ is the seed value of the cosmological constant, $\Lambda$ is its observed value (see \eqref{lambda0->Lambda}).
Further, since $H=0$, we must conclude that $\xi=\mathrm{Const}=0$, therefore, the Universe at the singular point is Euclidean, and the total energy density is equal to zero.

However, if we use the autonomous subsystem of the dynamical system \eqref{dot_xi-dot_Phi_e=0} -- \eqref{dZ/dt_e=0}, excluding the first equation \eqref{dot_xi-dot_Phi_e=0} from it, we will not get from this subsystem any condition on the Hubble parameter, which we will find from \eqref{Surf_Einst1_0_e=0}, substituting into this equation the values $Z=0$ from \eqref{Eq_sing_points_1-3} and $\Phi$ from \eqref{Phi_sing}. Thus, we obtain correct results for the characteristics of singular points (see, for example, \cite{Ignat21_1} -- \cite{Ignat22_1}, \cite{Ignat21_TMP}). The above example indicates the fact that not any dynamic functions can be good enough for a qualitative analysis of a dynamic system; in some cases, their poor choice can lead to system degeneration.

\section{Modified system of dynamic equations}
\subsection{Transformation of a dynamical system to a non-degenerate form}
To get rid of the indicated disadvantage, we transform the dynamical system \eqref{dH/dt_0} -- \eqref{Surf_Einst1_0} to new variables, noting that the right-hand sides of these equations depend on $\xi(t)$ only through the function $\psi(t)$, and expressing the scale-invariant function $\xi(t)$ from the relation \eqref{psi(t)} in terms of a couple of other scale-invariant functions:
\begin{equation}\label{xi(psi)}
\xi=-\ln\biggl|\frac{e\psi\Phi}{\pi_0}\biggr|.
\end{equation}
Thus, instead of \eqref{dH/dt_0} -- \eqref{Surf_Einst1_0} we get a new system of equations (only the first equation of the system formally changes)
\begin{eqnarray}\label{Sys1-Sys3}
\dot{\psi}=\psi\biggl(H-\frac{Z}{\Phi}\biggr)\; (\equiv G_1),\; \dot{\Phi}=Z\;(\equiv G_3),\\
\label{Sys2}
\dot{H}=- \frac{Z^2}{2}-\frac{4}{3\pi}e_z^4\Phi^4\psi^3\sqrt{1+\psi^2}\;(\equiv G_2),\\
\label{Sys4}
\dot{Z}=-3HZ-m_s^2\Phi +\Phi^3\biggl(\alpha-\frac{4e^4}{\pi}F_1(\psi)\biggr)\;(\equiv G_4);
\end{eqnarray}
the first integral of the system \eqref{Surf_Einst1_0} does not change.

\subsection{Singular points of the modified system\label{Sing_Points_sub}}

Let us now find the coordinates of singular points. In this case, the equation $G_1=0$ has two solutions, the first: $Z=H\Phi$ and the second: $\psi=0$. It is easy to see that the first solution finally brings us back to the previous situation ($H=0$). Let us therefore turn to the second solution. Taking $Z=0$ into account, this solution turns the equation $G_2=0$ into an identity. Considering that according to \eqref{F_1} $F_1(0)=0$, we obtain from \eqref{Sys4} solutions for the coordinate of the singular point, which coincide with the solutions for the one-field vacuum model \eqref{Phi_sing}. Thus, the dynamic equations \eqref{Sys1-Sys3} --\eqref{Sys4} did not impose any restrictions on $H(t)$. This value, as in the case of the vacuum field model, we will obtain from the first integral \eqref{Surf_Einst1_0}, taking into account $F_2(0)=0$:
\begin{equation}\label{H_sing}
\begin{array}{ll}
\displaystyle H_0^\pm=\pm\sqrt{\frac{\Lambda}{3}}, & \Phi=\Phi_0;\\
\displaystyle H_\pm=\pm\sqrt{\frac{\Lambda_0}{3}}, & \Phi=\Phi_\pm.\\
\end{array}
\end{equation}
Let us write out the coordinates of all six singular points of the system (the signs take on independent values):
\begin{eqnarray}\label{M_0}
M^\pm_0=\biggl[0,\pm\sqrt{\frac{\Lambda}{3}},0,0\biggr];\\
\label{M_pm}
M^\pm_\pm=\biggl[0,\pm\sqrt{\frac{\Lambda_0}{3}},\pm\frac{m_s}{\sqrt{\alpha}},0\biggr].
\end{eqnarray}

Note, firstly, that the coordinates of the singular points $[H,\Phi,Z]$ for the dynamical system \eqref{Sys1-Sys3} --\eqref{Sys4} with the integral condition \eqref{Surf_Einst1_0} coincide with the coordinates of the singular points of the vacuum-field model for the scalar singlet (see \cite{Ignat21_TMP}), which explains the previously noted features of the behavior of the cosmological model with charged trusses ions near singular points of the vacuum field model. Secondly, these points correspond to the zero value of the function $\psi(t)$, ($\xi\to+\infty$), i.e., to the late stages of cosmological evolution, at which the role of matter is negligible. Thirdly, we note that all similarity laws, together with the laws of scaling transformation of eigenvalues, are also preserved for the modernized dynamical system.

\subsection{Characteristic matrix and eigenvalues of a nondegenerate dynamical system}
Introducing an ordered set of new phase coordinates
\[ [\psi,H,\Phi,Z],\]
write the matrix of the dynamical system \eqref{Sys1-Sys3} --\eqref{Sys4} at singular points
\[ M_0=[0,H,\Phi,0] \Rightarrow
\mathbf{Y}=\left(\begin{array}{c}
0 \\ H \\ \Phi \\ 0\\
\end{array}
\right),\]
\begin{eqnarray}\label{B}
\mathbf{B}(M_0)=\left|\left|B_i^k \right|\right|\equiv \left|\left|\frac{\partial G_i}{\partial y_k}\right|\right|=
\left(\begin{array}{cccc}
H  & 0   &   0 & 0 \\
0  & 0 & 0 & 0 \\
 0 & 0 & 0 & 1 \\
\displaystyle 0 & 0 & -m_s^2+3\alpha\Phi^2 & - 3H \\
\end{array}\right),
\end{eqnarray}
where you need to substitute the values $\Phi$ and $H$ from \eqref{Phi_sing} and \eqref{H_sing}.

The eigenvalues of the matrix \eqref{B} are:
\begin{eqnarray}
\lambda_1=0,\; \lambda_2=H; \nonumber\\
\lambda_{3,4}=-\frac{3}{2}H\pm \frac{3}{2}\sqrt{H^2+\frac{4}{9}(3\alpha\Phi^2-m_s^2)}.\nonumber
\end{eqnarray}
Substituting here the values of $\Phi$ and $H$ from \eqref{Phi_sing} and \eqref{H_sing}, we finally find
eigenvalues of the matrix \eqref{B} at singular points:
\begin{eqnarray}\label{eigen_val_0}
\mathbf{M^\pm_0:}
\left\{\begin{array}{ll}
\lambda_2=& \pm\sqrt{\frac{\Lambda}{3}},\\[4pt]
\lambda_{3,4}=& \mp\frac{1}{2}\sqrt{3\Lambda} \pm \frac{1}{2}\sqrt{3\Lambda-4m^2_s};\\
\end{array}\right.
\end{eqnarray}
\begin{eqnarray}\label{eigen_val_pm}
\mathbf{M^\pm_\pm:}
\left\{\begin{array}{ll}
\lambda_2=& \pm\sqrt{\frac{\Lambda_0}{3}},\\[4pt]
\lambda_{3,4}=& \mp\frac{1}{2}\sqrt{3\Lambda_0} \pm \frac{1}{2}\sqrt{3\Lambda_0+8m^2_s}.\\
\end{array}\right.
\end{eqnarray}
Note that the expressions for the eigenvalues \eqref{eigen_val_0} -- \eqref{eigen_val_pm}, as well as the expressions for the coordinates of singular points \eqref{Phi_sing} and \eqref{H_sing}, according to the law of transformation of the fundamental parameters \eqref{trans_param} once again confirm the validity of the assertion \textbf{\ref{stat2}}.

\begin{stat}\label{stat3}$\blacksquare$
he coordinates of the eigenpoints of the dynamic system of the cosmological model $\Mod{M}$ at $H\not\equiv 0$ in the subspace $\mathbb{R}_3\equiv\{H,\Phi,Z\}\subset \mathbb{R}_4$, as well as the eigenvalues of the characteristic matrix, coincide with the coordinates of the eigenpoints and the eigenvalues of the characteristic matrix of the vacuum-field cosmological model. $\blacksquare$
\end{stat}

Note that the cited papers \cite{TMF_21}, \cite{Ignat21_1} -- \cite{Ignat22_1} did not reveal these singular points due to the choice of dynamical variables $[\xi,H,\Phi,Z]$, which leads to a degenerate characteristic matrix of the dynamical system. In these papers, the singular points of only one branch of the solutions of the equation $G_1=0$ in the \ref{Sing_Points_sub} section, corresponding to the infinite future of the Universe ($\xi\to+\infty,H\to0$), were identified and studied. Note also that the solution branch found here corresponds to $\psi\to0$, i.e., formally also corresponds to the $\xi\to+\infty$ case, however, the model remains in the inflation mode, i.e., the dominance of the scalar field over particles.

\section{Examples of numerical simulation of a dynamic\newline system}
So, according to the statements \textbf{\ref{stat1}} and \textbf{\ref{stat3}}, the coordinates of singular points \eqref{M_0} -- \eqref{M_pm} of the dynamical system under study are transformed under the $\mathcal{S}_k$ similarity transformation as follows
\begin{eqnarray}\label{M_0}
\tilde{M}^\pm_0=\biggl[0,\pm\frac{1}{k}\sqrt{\frac{\Lambda}{3}},0,0\biggr];\\
\label{M_pm}
\tilde{M}^\pm_\pm=\biggl[0,\pm\frac{1}{k}\sqrt{\frac{\Lambda_0}{3}},\pm\frac{m_s}{\sqrt{\alpha}},0\biggr],
\end{eqnarray}
\subsection{Standard Example}
Consider an example of numerical simulation of two similar systems $\Mod{M}$ and $\ModTilde{M}=\mathcal{S}_k(\Mod{M})$ with similarity coefficient $k=10^{4}$:
\begin{eqnarray}
\label{Mod0}
\Mod{M}: \Mod{P}=\bigl[\bigl[1,1,1,0.1\bigr]],3\cdot10^{-3}\bigr]; & \Mod{I}=[1,0,1];\\
\label{tilde_Mod0}
\ModTilde{M}: \ModTilde{P}=\bigl[\bigl[10^{-8},10^{-4},10^{-2},10^{-3}\bigr],3\cdot 10^{-11}\bigr]; & \ModTilde{I}=[1,0,1].
\end{eqnarray}

The eigenpoints of these models in the subspaces $\mathbb{S}_3=\{H,\Phi,Z\}$ and $\tilde{\mathbb{S}}_3=\{\tilde{H},\Phi,\tilde{Z}\}$ have the following coordinates (we derived approximate values for simplicity):
\begin{eqnarray}\label{Sing_Points_M}
\Mod{M}: \begin{array}{ll}
M_0=& \bigl[3.16\cdot10^{-2}, 0, 0\bigl];\\
M_\pm=& \bigl[2.90\cdot 10^{-1}, 1,0\bigl]\\
\end{array}\\
\label{Sing_Points_M_tilde}
\ModTilde{M}: \begin{array}{ll}
\tilde{M}_0=& \bigl[3.16\cdot10^{-6}, 0, 0 \bigl];\\
\tilde{M}_\pm=& \bigl[2.90\cdot 10^{-5}, 1, 0 \bigl]\\
\end{array}
\end{eqnarray}
As it is easy to see, the coordinates of singular points are transformed in exact accordance with the similarity laws \eqref{Sol_tilde} with the similarity coefficient $k=10^{4}$.

It is easy to calculate the eigenvalues of the characteristic matrix at these points (signs take independent values)
\begin{eqnarray}\label{eig_val_M}
\Mod{M}:& \begin{array}{ll}
k_0=& \mp 0.0474\mp i;\\
k_\pm=& \mp 1.915 \pm 1.044 \\
\end{array}\\
\label{eig_val_M_tilde}
\ModTilde{M}: & \begin{array}{ll}
\tilde{k}_0=& \mp 4.74\cdot10^{-6}\mp i\cdot10^{-4};\\
\tilde{k}_\pm=& \mp 1.915\cdot10^{-4},\ \pm 1.044\cdot10^{-4}\\
\end{array}
\end{eqnarray}
are the eigenvalues of the characteristic matrices of two similar dynamical cosmological systems are also transformed in exact correspondence with the similarity law \eqref{tilde_lambda=} with the similarity coefficient $k=10^{4}$. Thus, in the considered example, the points $M_0$ and $\tilde{M}_0$ are attracting foci, and the points $M_\pm$ and $\tilde{M}_\pm$ are saddle or nodal foci.

\TwoFigs{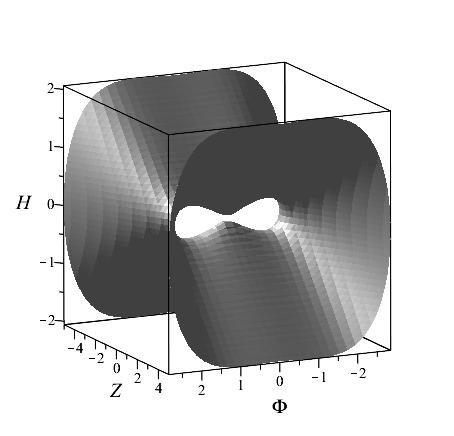}{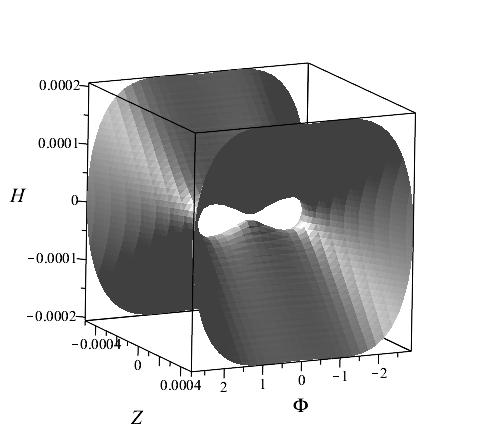}{\label{ignatev1}Projection of the Einstein-Higgs hypersurface of the $\ModTilde{M}$ \eqref{Mod0} model onto the hyperplane $\xi=0.1$.}{\label{ignatev2}Projection of the Einstein-Higgs hypersurface of the $\ModTilde{M}$ \eqref{tilde_Mod0} model onto the hyperplane $\tilde{\xi}=0.1$.}

On Fig. \ref{ignatev1} and \ref{ignatev2} show projections of the Einstein-Higgs surface of two models onto the hyperplane $\xi=\tilde{\xi}=0.1$ $S_3=[H,\Phi,Z,]$, and Fig. \ref{ignatev3} and \ref{ignatev4} are projections in the hyperplanes $Z=0.1$ and $\tilde{Z}=10^{-5}$ of the phase space of the dynamical system. As is easy to see from these figures, the graphs of the Einstein surfaces of two similar models are also similar in all projections - the scales along the $OH$ and $OZ$ axes are compressed by $10^4$ times for a similar model, the scales along the $O\xi$ and $O\Phi$ axes are preserved. Note also that according to the laws of similarity \eqref{Sol_tilde} the equation of the secant hyperplane $Z=\mathrm{Const}$ is also transformed.

\TwoFigs{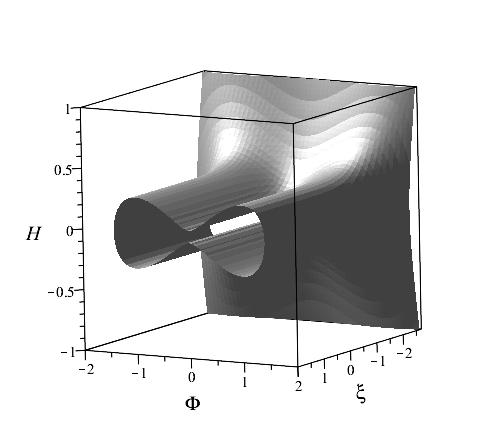}{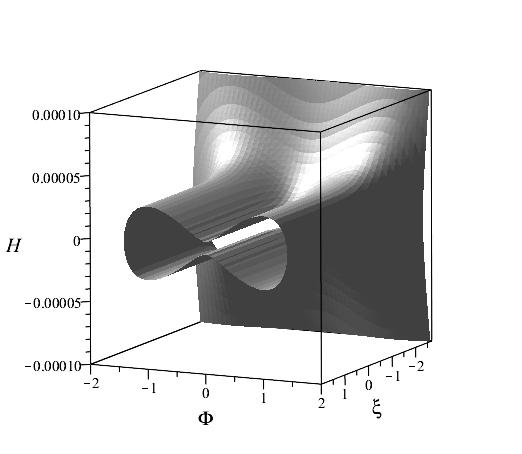}{\label{ignatev3}Projection of the Einstein-Higgs hypersurface of the $\Mod{M}$ \eqref{Mod0} model onto the hyperplane $Z=0.1$.}{\label{ignatev4}Projection of the Einstein-Higgs hypersurface of the $\ModTilde{M}$ \eqref{tilde_Mod0} model onto the hyperplane $\tilde{Z}=10^{-5}$.}

Next, in Fig. \ref{ignatev5} and \ref{ignatev6} are graphs of the evolution of the scale functions $\xi(t)$ and $H(t)$ in the $\Mod{M}$ and $\ModTilde{M}$ models, and Fig. \ref{ignatev7} and \ref{ignatev8} -- phase diagrams of the corresponding models. It can be seen that all corresponding pairs of graphs are similar with similarity coefficient $k=10^{4}$ in accordance with the similarity laws \eqref{Sol_tilde}.

On the charts in Fig. \ref{ignatev5} and \ref{ignatev6} in particular, one can observe strict time scale scaling: $\tilde{t}=10^4 t$.

\TwoFigs{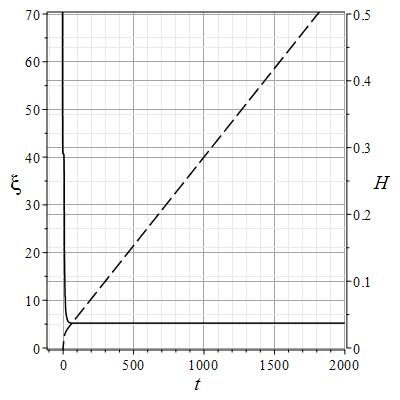}{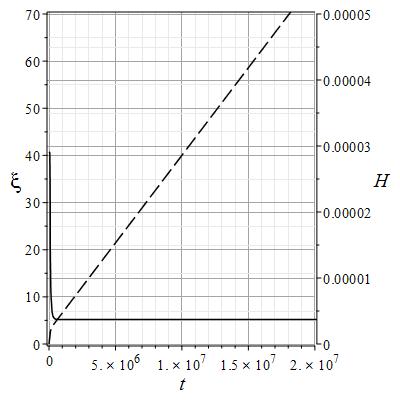}{\label{ignatev5}The evolution of scale functions $\xi(t)$ is a dashed line and $H(t)$ is a solid line in the model $\Mod{M}$ \eqref{Mod0}.}{\label{ignatev6}The evolution of the scale functions $\xi(t)$ is a dashed line and $H(t)$ is a solid line in the model $\ModTilde{M}$ \eqref{tilde_Mod0}.}

Thus, the results of numerical simulation strictly and clearly confirm all the similarity properties formulated above \textbf{\ref{stat1}} -- \textbf{\ref{stat3}} for similar models $\Mod{M}$ \eqref{Mod0} and $\ModTilde{M}$ \eqref{tilde_Mod0}.

\TwoFigs{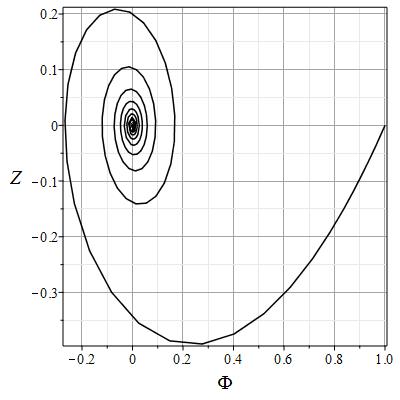}{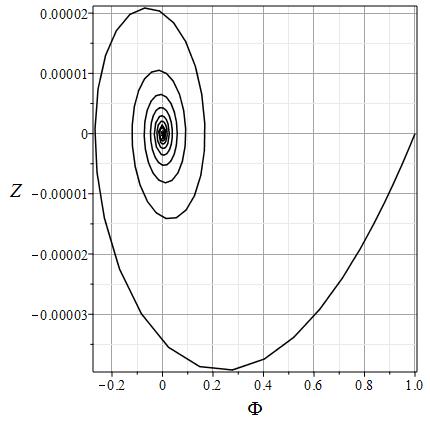}{\label{ignatev7}Phase diagram of the $\Mod{M}$ \eqref{Mod0} model in the $\{\Phi,Z\}$ plane.}{\label{ignatev8}Phase diagram of the $\ModTilde{M}$ \eqref{tilde_Mod0} model in the $\{\Phi,Z\}$ plane.}
\subsection{Possible violation of symmetry of similarity near saddle singular points}
Despite the fact that the results of the numerical simulation given above for the standard example reveal the strict fulfillment of the formulated similarity properties \textbf{\ref{stat1}} -- \textbf{\ref{stat3}}, cases of violation of the similarity symmetry of cosmological models when building numerical models are possible. These cases can take place at very small values of the scalar charge $e$, if in this case the phase trajectory of the model passes through an unstable singular point of the dynamical system. Consider this example:
\begin{eqnarray}
\label{Mod1}
\Mod{M}_\mathbf{1}: \Mod{P}=\bigl[\bigl[1,1,10^{-7},0.1\bigr]],3\cdot10^{-3}\bigr]; & \Mod{I}=[1,0,1];\\
\label{tilde_Mod1}
\ModTilde{M}_\mathbf{1}: \ModTilde{P}=\bigl[\bigl[10^{-8},10^{-4},10^{-9},10^{-3}\bigr],3\cdot 10^{-11}\bigr]; & \ModTilde{I}=[1,0,1].
\end{eqnarray}
Note that in this case the coordinates of the singular points coincide with the coordinates of the singular points of the \eqref{Sing_Points_M} and \eqref{Sing_Points_M_tilde} models considered above, since the parameters $e$ and $\pi_0$ do not affect these coordinates, as well as the nature of the points. Note also that the initial conditions $\Phi_0,Z_0$ in the above example, as in this case, coincide with the coordinates of the singular unstable points $M_\pm$ and $\tilde{M}_\pm$.
But in this case we reduced the scalar charge by $10^7$ times. On Fig. \ref{ignatev9} and \ref{ignatev10} plots of the evolution of the scale functions $\xi(t)$ and $H(t)$ in the $\Mod{M}_\mathbf{1}$ and $\ModTilde{M}_\mathbf{1}$ models are presented.
\TwoFigs{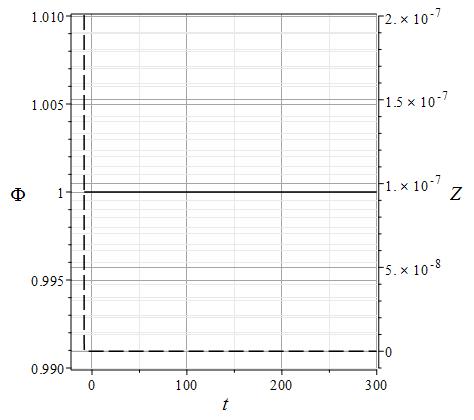}{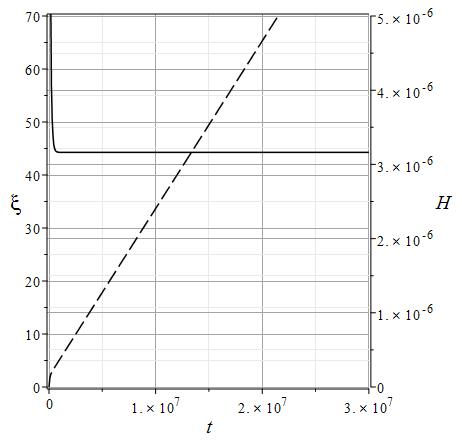}{\label{ignatev9}The evolution of the scaling functions $\xi(t)$ is a dashed line and $H(t)$ is a solid line in the $\Mod{M}_\mathbf{1}$ \eqref{Mod1} model.}{\label{ignatev10}The evolution of the scaling functions $\xi(t)$ is a dashed line and $H(t)$ is a solid line in the $\ModTilde{M}_\mathbf{1}$ \eqref{tilde_Mod1} model.}
These drawings, despite their outward similarity, demonstrate, however, a violation of the similarity symmetry. Indeed, the similarity coefficient for these models is equal to $k=10^4$. Therefore, according to the graph in Fig. \ref{ignatev9} the value $\xi=70$ should be reached in the model $\ModTilde{M}_\mathbf{1}$ at time $t\approx 2.4\cdot 10^6$, but from the graph in Fig. \ref{ignatev10} it can be seen that this value is reached at $t\approx 2\cdot 10^7$, i.e., an order of magnitude later. In this case, the value of the Hubble parameter according to the graph in Fig. \ref{ignatev9} in the $\ModTilde{M}_\mathbf{1}$ model should be about $H\approx 3\cdot10^{-5}$, but from the graph in Fig. \ref{ignatev10} we find the value $H\approx 3\cdot10^{-6}$, i.e., an order of magnitude smaller. To understand the situation, we present the phase diagrams of the models in the $\{\Phi,Z\}$ plane (Fig. \ref{ignatev11}, \ref{ignatev12}).
\TwoFigs{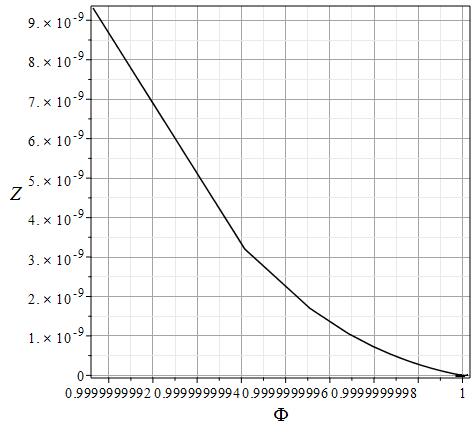}{ignatev8.jpg}{\label{ignatev11}Phase diagram of the $\Mod{M}_\mathbf{1}$ \eqref{Mod1} model in the $\{\Phi,Z\}$ plane.}{\label{ignatev12}Phase diagram of the $\ModTilde{M}_\mathbf{1}$ \eqref{tilde_Mod1} model in the $\{\Phi,Z\}$ plane.}
Commenting on the graph in Fig. \ref{ignatev11}, note that in fact, taking into account the accuracy of the calculations, this graph represents one point on the plane $\{\Phi,Z\}$: $\Phi=1,Z=0$, i.e., describes ``sticking'' of the trajectory at a singular point.

To demonstrate the influence of the singular point on phase trajectories, consider an example with initial conditions close to this point, replacing the initial value $\Phi_0=1$ with $\Phi_0=0.999$ close to it:
\begin{eqnarray}
\label{Mod1a}
\Mod{M}_\mathbf{1a}: \Mod{P}=\bigl[\bigl[1,1,10^{-7},0.1\bigr]],3\cdot10^{-3}\bigr]; & \Mod{I}=[0.999,0,1];\\
\label{tilde_Mod1a}
\ModTilde{M}_\mathbf{1a}: \ModTilde{P}=\bigl[\bigl[10^{-8},10^{-4},10^{-9},10^{-3}\bigr],3\cdot 10^{-11}\bigr]; & \ModTilde{I}=[0.999,0,1].
\end{eqnarray}
\TwoFigs{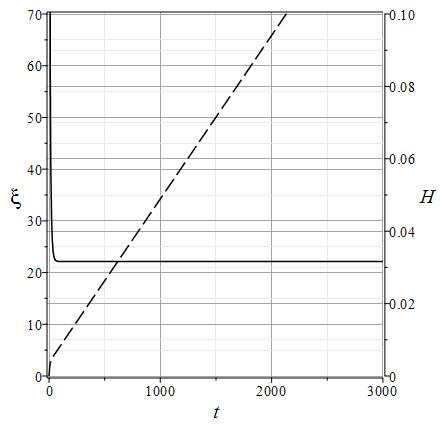}{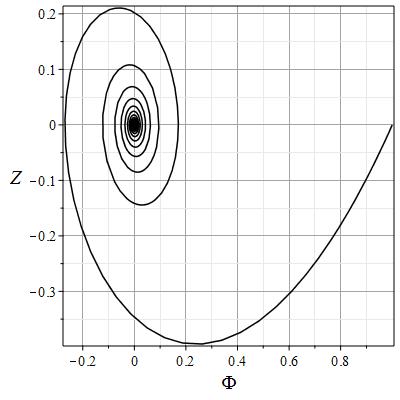}{\label{ignatev13}The evolution of the scaling functions $\xi(t)$ is a dashed line and $H(t)$ is a solid line in the $\Mod{M}_\mathbf{{1a}}$ \eqref{Mod1a} model.}{\label{ignatev14}Phase diagram of the $\Mod{M}_\mathbf{1a}$ \eqref{Mod1} model in the $\{\Phi,Z\}$ plane.}
On Fig. \ref{ignatev13} and \ref{ignatev14} plots of the evolution of scale functions $\xi(t)$ and $H(t)$ and phase diagrams in the $\{\Phi,Z\}$ plane of the $\Mod{M}_\mathbf{1a}$ model, plots of the $\ModTilde{M}_\mathbf{1a}$ model
in this case do not differ from the corresponding graphs in Fig. \ref{ignatev10} and \ref{ignatev12}). Thus, one can be convinced that the symmetry of similarity is restored with a slight shift of the initial conditions from the coordinates of the singular point. Nevertheless, the considered case shows that near the singular points of the dynamical application of the similarity transformation must be applied with caution.

\section*{Conclusion}

Concluding the article, we note, firstly, that the values of fundamental parameters (model $\tilde{\mathbf{M}}$)\footnote{Such values correspond, for example, to the scales of field models like SU(5).} are physically realizable:
\begin{eqnarray}\label{Phys_Par}
\alpha\lesssim10^{-8}; m\lesssim10^{-4};e\lesssim10^{-2}.
\end{eqnarray}
But for such small values of the fundamental parameters, purely technical difficulties in the numerical integration of a nonlinear system of dynamic equations make it possible to extend calculations only up to time values of the order of $t\lesssim 10^4$ (see, for example, \cite{TMF_21}, \cite{Ignat21_1}). However, performing a scaling transformation with a similarity coefficient of the order of $k=10^4$, we will pass to the cosmological model $\mathbf{M}$ with parameters $\alpha=m=e=1$, which is already amenable to numerical simulation up to much larger time values $t\sim 10^8$. Using the results of integrating the $\mathbf{M}$ model according to the scaling transformation rules, we thereby extend the results for the original $\tilde{\mathbf{M}}$ model up to times of the order of $t\gtrsim 10^8$.

Secondly, a physically important circumstance is that when the $\Mod{M}$  and $\ModTilde{M}$ cosmological model is transmitted from the studied cosmological model $\tilde{M}$ with the cene of the sake of $k\sim 10^4\div10^5$, we stretch the temporary interval of cosmological evolution, moving to the time, moving to the time, moving to the time $t\gg t_{pl}$, in which a quantum-field consideration of the cosmological model is not required, but a fairly classic description. Making the reverse transition from the classical model $\ModTilde{M}$ to the similar model $\Mod{M}$ with the similarity coefficient $k^{-1}$, we obtain the classical model at times comparable to the Planck ones. This model, however, by no means claims any physical meaning. It serves only as a \emph{convenient computational model} similar to the classical cosmological model under study for large evolution times $t\gg t_{pl}$.

Thirdly, the explicit dependence of the solutions \eqref{Sol_tilde} of the dynamic system of equations describing the cosmological model on the similarity coefficient makes it possible to analytically extend the obtained numerical solutions to similar models and similar time intervals, which greatly expands the possibilities of analyzing numerical models. At the same time, the results of qualitative analysis describing the global properties of dynamic models are also being disseminated.

Finally, fourthly, it is obvious that the established laws of similarity of dynamical systems can be successfully applied to other cosmological models with scalar fields, which is an important circumstance. In particular, the previously considered cosmological models based on scalar multipoles, for example, an asymmetric scalar doublet \cite{Ignat21_TMP} or a multifield model with an exponential interaction potential \cite{Leon18}.

\subsection*{Funding}
This paper has been supported by the Kazan Federal University Strategic Academic Leadership Program.


\begin{thebibliography}{30}
%
\bibitem{Klain}
S. J. Kline, \emph{Similarity and approximate methods}, Moskow: Mir (1968).

\bibitem{Sedov}
L. I. Sedov, \emph{Methods of similarity and dimension in mechanics}, Moskow: Nauka (1977); (1987).

\bibitem{Dibai}
E. A. Dibay, S. A. Kaplan, \emph{Dimensions and similarity of astrophysical quantities}, Moskow: Nauka (1976).

\bibitem{TMF_21}
Yu.G. Ignat'ev and D.Yu. Ignat'ev, ``Cosmological Models Based on a Statistical System of Scalar Charged Degenerate Fermions and an Asymmetric Higgs Scalar Doublet'', \emph{Theoret. and Math. Phys.}, \textbf{209}  1437-1472 (2021) (2021); arXiv:2111.00492 [gr-qc].

\bibitem{Ignat21_1}
Yu.G. Ignat'ev, A.A. Agathonov and D.Yu. Ignatyev,``Cosmological evolution of a statistical system of degenerate scalar-charged fermions with an asymmetric scalar doublet. I. Two-component system of assorted charges.'',  \emph{Gravit. Cosmol}., \textbf{27:4} pp. 338-349 (2021); arXiv:2203.11946 [gr-qc].

\bibitem{Ignat22_1}
Yu.G. Ignat'ev, A.A. Agathonov and D.Yu. Ignatyev, ``Cosmological Evolution of a Statistical System of Degenerate Scalarly Charged Fermions with an Asymmetric Scalar Doublet.
II. One-Component System of Doubly Charged Fermions.'',	\emph{Gravit. Cosmol}.,  \textbf{28:1},  10-24 (2022);  arXiv:2203.12766 [gr-qc].

\bibitem{Yu_Unst_GC22_1}
Yu. G. Ignat'ev, ``Gravitational-Scalar Instability of a Two-Component Degenerate System of Scalarly Charged Fermions with Asymmetric Higgs Interaction'',	\emph{Gravit. Cosmol}., \textbf{28}, 25 (2022); arXiv:2203.11948 [gr-qc].

\bibitem{Yu_Unst_GC22_2}
Yu. G. Ignat'ev, ``Single-Field Model of Gravitational-Scalar Instability. I. Evolution of Perturbations'',	\emph{Gravit. Cosmol}., \textbf{28}, 275 (2022); arXiv:2207.05066 [gr-qc].

\bibitem{Yu_Unst_GC22_3}
Yu. G. Ignat'ev, ``Single-Field Model of Gravitational-Scalar Instability. II. Black Hole Formation'',	\emph{Gravit. Cosmol}. \textbf{28}, 375 (2022); arXiv:2211.14507v1 [gr-qc].

\bibitem{Yu_Unst_GC23_1}
Yu. G. Ignat'ev, ``Two-Field Model of Gravitational-Scalar Instability and Formation of Supermassive Black Holes in the Early Universe'',	\emph{Gravit. Cosmol}. \textbf{28}, 163 (2023); arXiv:2305.15456 [physics.gen-ph].

\bibitem{YuTMF_23}
Yu. G. Ignat'ev, ``Evolution of spherical perturbations in the cosmological environment of degenerate scalarly charged fermions with the Higgs scalar interaction'', \emph{Theoret. and Math. Phys.}, \textbf{215:3} 862-892 (2023); arXiv:2306.17185 [gr-qc].

%
\bibitem{Ignat21_TMP}
Yu.G. Ignat'ev, I.A. Kokh, ``Complete cosmological model based on a asymmetric scalar Higgs doublet'', \emph{Theoret. and Math. Phys}. \textbf{207}, 514 (2021); arXiv:2104.01054  
%

\bibitem{Yu_Dima_20GC2}
Yu. G. Ignat'ev and D. Yu. Ignat'ev, Gravit. Cosmol. \textbf{26}, 249 (2020); arXiv:2007.04392 [gr-qc].


\bibitem{Bogoyav}
O. I. Bogoyavlensky, \emph{Methods of the qualitative theory of dynamical systems in astrophysics and gas dynamics}, Moskow: Nauka (1980).

%
\bibitem{Leon18}
Genly Leon (Catolica del Norte U.), Andronikos Paliathanasis, Jorge Luis Morales, ``The past and future dynamics of quintom dark energy models'', \emph{Eur. Phys. J}. \textbf{C 78}, 753 (2018).
%
%
\end{thebibliography}
\end{document}